\documentclass[pre,aps,twocolumn,nofootinbib]{revtex4-1}
\usepackage[latin1]{inputenc}
\usepackage[english]{babel}
\usepackage{graphicx}
\usepackage{color}
\usepackage{amsmath}
\usepackage{amssymb}
\usepackage{epstopdf}
\usepackage{multirow}
\usepackage{caption}
\usepackage{amsmath}
\usepackage{amssymb}
\begin{document}

\title{\bf Simple regular black hole with logarithmic entropy correction}
\author{Nicol\'as Morales--Dur\'an $^{1}$, Andr\'es F. Vargas $^{1}$, Paulina Hoyos--Restrepo and Pedro Bargue\~no}
\affiliation{Departamento de F\'{\i}sica,
Universidad de los Andes, Apartado A\'ereo {\it 4976}, Bogot\'a, Distrito Capital, Colombia (p.bargueno@uniandes.edu.co)
\\ $^{1}$Shared first authorship
}
\begin{abstract}
A simple regular black hole solution satisfying the weak energy condition is obtained within Einstein--non--linear electrodynamics theory.
We have computed the thermodynamic properties of this black hole by a careful analysis of the horizons and we have found that the
usual Bekenstein--Hawking entropy gets corrected by a logarithmic term. Therefore, in this sense our model realizes some quantum gravity predictions
which add this kind of correction to the black hole entropy. In particular, we have established some similitudes between our model and a quadratic generalized
uncertainty principle. This similitude has been confirmed by the existence of a remnant, which prevents complete evaporation, in agreement 
with the quadratic generalized uncertainty principle case.

\end{abstract}

\maketitle

\section{Introduction}
Spacetime singularities are one of the most problematic features of general relativity. 
Physics breaks down there and unpredictability appears to be unavoidable. Among all the predictions of general relativity, black holes (BHs) are
usually considered one of the most fascinating objects which populate our universe, and are frequently used to test different attempts to unify
general relativity with quantum mechanics. After the singularity theorems by Hawking and Penrose \cite{Penrose1965,book} 
(an excellent overview of these theorems and subsequent extensions can be found in \cite{Senovilla}), BHs are known to have a singularity inside them.

These theorems can be circumvented and regular BHs, that is, solutions of Einstein equations that have horizons but are regular everywhere, can be constructed.
In particular, charged regular BH solutions exist in the framework of Einstein--nonlinear electrodynamics (NLED) theory.
The interest in these theories is twofold. First, quantum corrections to Maxwell theory can be described by means of non--linear effective Lagrangians that define NLEDs as, for example,
the Euler--Heisenberg Lagrangian \cite{HE,Sch}, which is effectively described by  Born--Infeld (BI) theory \cite{BI}. Even more, higher order corrections 
give place to a sequence of effective Lagrangians which are polynomials in the field invariants \cite{Bia}.
And second, in case of dealing with open bosonic strings, the resulting tree--level
effective Lagrangian is shown to coincide with the BI Lagrangian \cite{PLB1985,NPB1997}.
These NLED theories, when coupled to gravity, give place to very interesting phenomena as, for instance, the appereance of generalized 
Reissner--Nordstr\"om geometries in the form of BI--like solutions \cite{Garcia1984,Breton2003,Hendi2013}. Interestingly, exact regular BH geometries in 
the presence of NLED were obtained in \cite{Ayon1998,Ayon1999a,Ayon1999b,Ayon2000,Ayon2005,Dym2004}. In particular, the Ay\'on--Beato and
Garc\'{\i}a solution \cite{Ayon1998}, further discussed in \cite{Baldo2000,Bronni2001}, extended the preliminary attempt of Bardeen \cite{Bardeen1968} 
to obtain regular BH geometries. Moreover, BHs with the Euler--Heisenberg 
effective Lagrangian as a source term were examined in \cite{Yajima2001}, and a similar type of solutions with Lagrangian densities that are 
powers of Maxwell's Lagrangian were analyzed in \cite{Hassaine2008}.
 
The plausibility of these solutions is usually checked with the help of 
energy conditions. In fact, if a BH is regular, the strong enery condition is violated somewhere inside the horizon \cite{Zas2010} but the weak or 
dominant energy conditions could be satisfied everywhere \cite{Dym2004}. Moreover, as pointed out in \cite{Ansoldi2007}, regular BHs that satisfy the 
weak energy condition (WEC) and their energy--momentum tensor is such that $T^{0}_{\,\;0}=T^{1}_{\,\;1}$ have a de Sitter behaviour at $r\rightarrow 0$. 
Regular BH solutions possessing this symmetry, some of them satisfying the WEC and 
with an asymptotically Reissner--N\"ordstrom behaviour, have been constructed in the framework of Einstein--NLEDs
\cite{Balart2014a,Balart2014b}. In a recent work \cite{Contreras2016},  several black hole metrics corresponding to nonlinearly charged 
black holes which were shown to be consistent with a logarithmic correction to the Bekenstein--Hawking entropy formula were constructed. 
The main drawback of this work was that the WEC was shown to be perturbatively violated at order $q^2$.  Therefore, as stated
in \cite{Contreras2016},  we think that it would be interesting to investigate whether or not is possible
to obtain effective regular BH geometries with reproduce the logarithmic correction without violating this energy condition.

In this work we tackle this problem and construct a new and very simple static and spherically symmetric regular BH solution, 
obtained within Einstein--NLED theory. 
Our result will be based on a useful formula relating the electric field, which will be imposed to be Coulomb--like, 
with the curvature invariants $R^{\mu\nu}R_{\mu\nu}$ and $R$. This BH will be shown to be Reissner--N\"ordstrom--like at infinity. As stated before,
 the WEC will be shown to be satisfied everywhere. Moreover, after a careful analysis of the horizons, the entropy and heat capacity will reveal 
that our model realizes some quantum gravity predictions which add a logarithmic correction to the BH entropy and which give place to a remnant. 
Finally, some conclusions are established regarding a possible realization of a quadratic generalized uncertainty principle by NLED.

\section{Preliminaries}

In geometrized units, Einstein's equations ($\Lambda =0$) read
\begin{equation}
R_{\mu\nu}-\frac{1}{2}R\, g_{\mu\nu} = 8 \pi T_{\mu\nu},
\end{equation}
where $T_{\mu\nu}$ is the energy--momentum tensor.

Let us form the following curvature invariants:
\begin{eqnarray}
\mathrm{K}&=&R^{\mu\nu\rho\sigma}R_{\mu\nu\rho\sigma} \nonumber   \\
\mathrm{Ric}^2&=&R^{\mu\nu}R_{\mu\nu} \nonumber \\
\mathrm{R}&=& g^{\mu\nu}R_{\mu\nu}.
\end{eqnarray}
As pointed out in \cite{Cherubini2002} in the four dimensional case, 
the non--Weyl part of the curvature determined by the matter content can be separated by showing that
\begin{equation}
\label{main}
4 \mathrm{Ric}^{2}-\mathrm{R}^{2}=(16 \pi)^{2}\left(T^{\mu\nu}T_{\mu\nu}+\frac{3-p}{(2-p)^{2}}T^{2} \right) ,
\end{equation}
where $T=g^{\mu\nu}T_{\mu\nu}$ is the trace of the energy--momentum tensor and $p$ is the dimension of the spacetime.

For simplicity let us take spherically symmetric and static solutions  given by ($p=4$)
\begin{equation}
\label{metric}
ds^{2}=-f(r)dt^{2}+f(r)^{-1}dr^{2}+r^{2} d\Omega^{2}.
\end{equation}
For the matter content we choose certain NLED. Assuming that the corresponding Lagrangian only depends on one of the two field invariants,
a particular choice for an energy--momentum tensor for NLED is
\begin{equation}
\label{nlT}
T^{\mu\nu}=-\frac{1}{4\pi}\left[\mathcal{L}(F)g^{\mu\nu}+\mathcal{L}_{F}F^{\mu}_{\;\;\rho}F^{\rho \nu} \right]
\end{equation}
where $\mathcal{L}$ is the corresponding Lagrangian, $F=\frac{1}{4}F_{\mu\nu}F^{\mu\nu}$ and $\mathcal{L}_{F}=\frac{d\mathcal{L}}{dF}$.

On one hand, in the electrovacuum case, and considering only a radial electric field as the source, that is,
\begin{equation}
\label{eqmaxwell}
F_{\mu \nu}=E(r)\left(\delta^{t}_{\mu}\delta^{r}_{\nu}-\delta^{r}_{\mu}\delta^{t}_{\nu} \right),
\end{equation}
Maxwell equations read
\begin{equation}
\nabla_{\mu}\left(F^{\mu\nu}\mathcal{L}_{F}\right)=0,
\end{equation}
thus, 
\begin{equation}
\label{max}
E(r)=-\frac{q}{4\pi r^{2}}(\mathcal{L}_{F})^{-1}.
\end{equation}
On the other hand, the components of the Einstein tensor and the curvature invariants are given by
\begin{eqnarray}
\label{compe}
G_{tt}&=& -\frac{f(r) \left[-1+f(r)+r f'(r)\right]}{r^2} \nonumber \\
G_{rr}&=&  \frac{-1+f(r)+r f'(r)}{r^2 f(r)}\nonumber \\
G_{\theta \theta}&=& \frac{1}{2} r \left[2 f'(r)+r f''(r)\right] \nonumber \\
G_{\phi\phi}&=&\sin^{2}\theta\, G_{\theta \theta}.
\end{eqnarray}
and
\begin{eqnarray}
\label{inv}
&&R^{\mu\nu\rho\sigma}R_{\mu\nu\rho\sigma}=\frac{4 \left[(-1+f(r))^2+r^2 f'(r)^2\right]}{r^4}+f''(r)^2 \nonumber   \\
&&R^{\mu\nu}R_{\mu\nu}= \frac{4+4 f(r)^2+8 r^2 f'(r)^2+8 f(r) \left[-1+r f'(r)\right]}{2 r^4} \nonumber \\
&&+\frac{r^4 f''(r)^2+4 r f'(r) \left[-2+r^2 f''(r)\right]}{2 r^4} \nonumber \\
&&R=g^{\mu\nu}R_{\mu\nu}=-\frac{2 \left[-1+f(r)+2 r f'(r)\right]}{r^2}-f''(r),
\end{eqnarray}
respectively. 
\\
\\
Therefore, taking into account that
\begin{equation}
\label{identity}
r^{2} G_{tt}+f(r) G_{\theta\theta}=-8\pi r^{2} f(r) E(r)^{2}\mathcal{L}_{F}
\end{equation}
and using Eqs. (\ref{max}),  (\ref{compe}) and (\ref{inv}), 
we arrive to the following expression for the electric field:
\begin{eqnarray}
\label{electric}
E(r)&=&\frac{1}{4q}\left[r^{2}f''(r)+2\{1-f(r)\} \right] \nonumber \\
&=&\frac{r^{2}}{4q}\sqrt{4 R^{\mu\nu}R_{\mu\nu} -R^{2}}.
\end{eqnarray}
\\ \\
It is important to point out that Eqs. (\ref{identity}) and (\ref{electric}) are also valid for $\Lambda\ne0$, as can be easily shown by direct calculation.
Moreover, Eq. (\ref{electric}) follows directly from Eqs. (\ref{main}), (\ref{nlT}) and (\ref{eqmaxwell}).
\\ \\
As a simple example of the applicability of Eq. (\ref{electric}), let us consider the case of a Reissner--N\"ordstrom BH whose relevant curvature invariants are given by 
$R^{\mu\nu}R_{\mu\nu}=\nobreak 4 q^{4}/r^{8}$ and $R=0$. Therefore, $4 R^{\mu\nu}R_{\mu\nu}-\mathrm{R}^{2}=16 q^{4}/r^{8}$ and $E(r)=q/r^{2}$, as expected.

As a bypass of Eqs. (\ref{max}) and (\ref{electric}), 
the following ordinary differential equation for $f(r)$ can be derived:
\begin{equation}
4 R^{\mu\nu}R_{\mu\nu}-\mathrm{R}^{2} = \left(\frac{q^2}{\pi r^4 \mathcal{L}_{F}}\right)^{2}.
\end{equation}
In the Maxwell case ($\mathcal{L}_{F}=-1$), the solutions are given by $f(r)=1-\frac{C_{1}}{r}\pm \frac{q^2}{4 \pi r^2}+C_{2}r^{2}$. 
The solution with positive sign coincides with the Reissner--N\"ordstrom one if $C_{1}=2 M$ and $C_{2}=0$ (we think that this is a simple way of obtaining
this well known solution). The negative sign corresponds to the quasicharged Einstein--Rosen bridge \cite{Einstein1935}, 
which is usually assumed to be an {\it ad hoc} modification
of the theory corresponding to a negative electromagnetic energy density. This is a clear example of how the Einstein equations do not say anything
about energy conditions, which have to be imposed apart from the dynamics.
\\
\\

In the general case, taking $\mathcal{L}_{F}=g(r)$, we get
\begin{widetext}
\begin{equation}
\label{long}
f(r) = r^2 A_1+\frac{A_2}{r}\pm r^2 \int_1^r \frac{\frac{q^2}{\pi  g(y)}-2 y^2}{3 y^5} \, d y+
\frac{\int_1^r \left(\frac{2}{3}\mp\frac{q^2}{3 \pi  g(y) y^2}\right) \, d y}{r}, 
\end{equation}
\end{widetext}
where $A_{1}$ and $A_{2}$ are arbitrary constants. It can be seen that only the (first) positive sign choice gives place to the correct $q=0$ case. 
Within this choice, Eq. (\ref{long}) simplifies to
\begin{widetext}
\begin{equation}
\label{longbis}
f(r) = 1+\frac{C_1}{r}+C_2 r^{2}+
\frac{q^{2}}{3\pi}\left(r^{2} \int_1^r \frac{dy}{g(y)y^5}
\mp r^{-1}\int_1^r \frac{dy}{g(y) y^2}\right), 
\end{equation}
\end{widetext}

Therefore, if a NLED model depending on $\mathcal{L}(F)$ is given, then Eq. (\ref{long}) can be used to calculate the 
corresponding geometry, which solves Einstein equations.
We note that the opposite way is usually considered when regular BH solutions are looked for. That is, one starts from certain metric and tries
to look for some NLED model such that the coupled Einstein--non--linear theory is satisfied.

This underlying NLED theory can be obtained using the $P$ framework \cite{Salazar1987}, which is somehow dual to the $F$ framework. After introducing the
tensor $P_{\mu\nu}=\mathcal{L}_{F}F_{\mu\nu}$ together with its invariant $P=-\frac{1}{4}P_{\mu\nu}P^{\mu\nu}$,
one considers the Hamiltonian--like quantity
\begin{equation}
\label{H}
\mathcal{H}=2 F \mathcal{L}_{F} -\mathcal{L}
\end{equation}
as a function of $P$, which specifies the theory. Therefore, the Lagrangian can be written as a function of $P$
as
\begin{equation}
\label{L}
\mathcal{L}=2 P \frac{d\mathcal{H}}{d{P}}-\mathcal{H}. 
\end{equation}
Finally, by reformulating the coupled Einstein--NLED equations
in terms of $P$, $\mathcal{H}(P)$ is shown to be given by \cite{Bronnikov2001}
\begin{equation}
\mathcal{H}(P)=-\frac{1}{r^2}\frac{d m(r)}{dr},
\end{equation}
where the mass function $m(r)$
is such that $f(r)=1-\frac{2 m(r)}{r}$.

\section{A simple regular black hole solution}

Let us assume a NLED theory such that, when dealing only whith static and spherically symmetric
sources, the corresponding electric field is regular everywhere. Let us consider a simple ansatz for the electric field, which we express as
\begin{equation}
\label{imposed}
E(r)=\frac{q r^n}{\left(r+a\right)^{n+2}},
\end{equation}
where $a>0$ and $n \in \mathbb{R}^{+}$. From this choice it is clear that $\lim_{r\rightarrow \infty} E(r)\rightarrow q/r^2$.

As commented before, the corresponding geometry can be obtained with the help of Eq. (\ref{electric}).
%
\\
After a long but straightforward calculation we get (for simplicity only $n\in \mathbb{N}$ will be considered):\\
\begin{itemize}
\item $n=0$. In this case the first two curvature invariants diverge at $r=0$. $K$ can be made regular everywhere after choosing
$a=\nobreak -\frac{q^2}{3C_{2}}$.
\item $n=1$. In this case the three curvature invariants diverge at $r=0$.
\item $n=2$. In this case $K$ diverges at $r=0$.
\item $n=3$. In this case the corresponding geometry is given by (choosing the appropriate sign)
\begin{equation}
f(r)=1+C_{1}r^2+\frac{C_{2}}{r}+\frac{q^2 a}{\left(a+r \right)^3}\left(1+\frac{a}{3r}+\frac{r}{a} \right),
\end{equation}
where $C_{1,2}$ are arbitrary constants. Moreover, the curvature invariants are regular provided $a=\nobreak -\frac{q^2}{3C_{2}}$. Let us analize this case
in more detail.
\end{itemize}

After considering the Schwarzschild--de Sitter metric, we get that $C_{1}=\frac{\Lambda}{3}$ and $C_{2}=-2M$. Therefore, $a=\frac{q^2}{6M}$ and we can write
\begin{widetext}
\begin{equation}
\label{theeq}
f(r)=1-\frac{2M}{r}+\frac{\Lambda}{3}r^2+\frac{q^2}{6 M \left(\frac{q^{2}}{6 M }+ r\right)^3}\left(q^2+\frac{q^4}{18}+r\right),
\end{equation}
\end{widetext}
which is of the form of Eq. (\ref{longbis}). Moreover, this solution is asymptotically Reissner--N\"ordstrom--de Sitter, 
\begin{equation}
f(r)=1-\frac{2M}{r}+\frac{\Lambda}{3}r^2+\frac{q^2}{r^2}-\frac{q^4}{3 M r^3}+\mathcal{O}\left(\frac{1}{r}\right)^4
\end{equation}
and, when $r\rightarrow 0$,
the metric function behaves as de Sitter space as
\begin{equation}
\label{limit}
f(r)\rightarrow 1 + \left(\frac{\Lambda}{3}-\frac{432 M^4}{q^6}\right) r^2.
\end{equation}

The curvature invariants are given by
\begin{widetext}
\begin{eqnarray}
R&=&-4 \Lambda + \frac{5184 M^4 q^4}{\left(q^2+6 M r\right)^5} \nonumber \\
R^{\mu\nu}R_{\mu\nu}&=&36 \left(\frac{\Lambda}{3}^2-\frac{288 \Lambda M^4 q^4}
{\left(q^2+6 M r\right)^5}+\frac{186624 M^8 q^4 \left(q^4+36 M^2 r^2\right)}{\left(q^2+6 M r\right)^{10}}\right) \nonumber \\
K&=&24 \left(\frac{\Lambda}{3}^2-\frac{288 \Lambda M^4 q^4}{\left(q^2+6 M r\right)^5}+\frac{186624 M^8 \left(q^8+126 M^2 q^4 r^2-216 M^3 q^2 r^3+648 M^4 r^4\right)}{\left(q^2+6 M r\right)^{10}}\right).
\end{eqnarray}
\end{widetext}
Therefore, the spacetime is regular everywhere, as stated before. The corresponding electric field is given by
\begin{equation}
\label{field}
E(r)=\frac{7776 M^{5} q\, r^3}{\left(q^2+6 M r \right)^{5}}=\frac{q}{r^2}-\frac{5 q^3}{6 M r^3}+ \mathcal{O}\left(\frac{1}{r}\right)^{4}
\end{equation}
%
%
Note that, as shown in Eq. (\ref{field}), the electric field, which is regular everywhere, 
is given by a very simple expression, in contrast to other works (see, for example, Refs. 
\cite{Ayon1998,Ayon1999a,Ayon1999b,Ayon2000,Ayon2005,Dym2004,Balart2014a,Balart2014b}). 
Moreover, as in the case of a BI electric field, it develops a maximum, but this time not at $r=0$. 
This feature could be related with the fact that a BH with BI electric field is not regular (see, for example, Ref. \cite{Garcia1984}), in contrast with our 
case. In other words, it appears as if the BI field was shifted to make our solution regular everywhere. In this sense, the spirit of BI NLED theory is 
recovered.
 
%
%
\section{Non-linear electrodynamics for the regular black hole and energy conditions}
Up to this point, nothing has been said about the plausibility of the obtained solution, except that it seems to be related to the BI one. We can shed
more light on the corresponding matter content by computing the structural function $\mathcal{H}(P)$ for the NLED theory.\\\\
For our $n=3$ solution, the mass function is given by
\begin{equation}
m(r)=\frac{216 M^4 r^3}{\left(q^2+6 M r\right)^3}.
\end{equation}
We point out that this mass functions tends to $M$ both in the $q=0$ and in the asymptotic spatial limit. 
From this, the structural function as a function of $r$ is given by
\begin{equation}
\mathcal{H}(r)=-\frac{m^{\prime}(r)}{r^2}
\end{equation}
Now, $r$ and $P$ are related by means of
\begin{equation}
r=\left(\frac{q^2}{2P}\right)^{\frac{1}{4}}
\end{equation}
Then we obtain
\begin{equation}
\mathcal{H}(P)=-\frac{648M^4q^2}{\left(q^2+2^{3/4}\cdot3M\left(\frac{q^2}{P}\right)^{1/4}\right)^4}
\end{equation}
It is interesting to perform a series expansion of the structural function around $P=0$ to compare our electrodynamics with other previous results. After 
introducing the parameter $s=\frac{q}{2M}$, we have
\begin{widetext}
\begin{equation}
\label{ourH}
\mathcal{H}(P)=-P+\frac{2^{1/2}\sqrt q sP^{\frac{5}{4}}}{3}-\frac{10\sqrt2 qs^2P^{\frac{3}{2}}}{9}+\frac{2^{3/4}\cdot20q^{3/2}s^3P^{\frac{7}{4}}}{27}-\frac{70q^2s^4P^2}{81}+\mathcal{O}(P)^{9/4}.
\end{equation}
\end{widetext}
A couple of comments are in order here. First of all, let us note that Maxwell's theory, $\mathcal{H}(P)=-P$, is recovered for small fields. In
addition, a quadratic BI--like term appears (fifth term in the RHS of Eq. (\ref{ourH})). This quadratic term is easy to interpret in light
of the cutoff field which is an essential ingredient of BI--theory. Therefore, the difficulty of interpreting this NLED theory can
be adscribed to the other terms which appear in the RHS of Eq. (\ref{ourH}). In spite of this, let us note that similar terms have appeared since
the discovery of the first exact regular BH solution by Ay\'on--Beato and Garc\'{\i}a \cite{Ayon1998}. In fact, the Hamiltonian function presented in
Ref. \cite{Ayon1998} can be expanded for weak fields to give
\begin{widetext}
\begin{equation}
\mathcal{H}_{AB}(P)=-P+\frac{3\cdot  2^{1/4} \sqrt{q} P^{5/4}}{s}-6 \sqrt{2 q^2} P^{3/2}+\frac{15 q^{3/2} P^{7/4}}{2^{1/4} s}-30 q^2 P^2 +\mathcal{O}(P)^{9/4},
\end{equation}
\end{widetext}
which except for some constants coincide with our Eq. (\ref{ourH}).

In spite of some drawback related with the interpretation of some terms which appear in Eq. (\ref{ourH}), the NLED we are considering has a nice feature because
the WEC is satisfied. In particular, with the help of $m(r)$, this WEC can be expressed as \cite{Balart2014a}
\begin{eqnarray}
&&\frac{m'(r)}{r^2}\ge 0 \nonumber \\
&&\frac{2 m'(r)}{r}-m''(r) \ge 0,
\end{eqnarray}
where the prime denotes derivative with respect to $r$. 

Therefore, using Eq. (\ref{theeq}) whith $\Lambda=0 $ we get
\begin{eqnarray}
&&\frac{m'(r)}{r^2}=\frac{648 M^4 q^2}{\left(q^2+6 M r\right)^4}\nonumber \\
&&\frac{2 m'(r)}{r}-m''(r) =\frac{15552 M^5 q^2 r^2}{\left(q^2+6 M r\right)^5}
\end{eqnarray}
\\
and, therefore, the WEC is satisfied everywhere (as commented in the introduction, given the regularity of the BH solution, the strong enery condition is 
violated somewhere inside the horizon \cite{Dym2004}. However, we left this calculation to the interested reader because we think it does not contribute 
to the subsequent discussion). Note that this agrees with the existence of the de Sitter behaviour \cite{Ansoldi2007}
for the metric function when $r\rightarrow 0$, given by Eq. (\ref{limit}).

\section{Event horizons}

In order to study the existence of event horizons in this geometry we have to find the values of $r$ where $g^{rr}=0$ - i.e: the set of values of $r$ for which $f(r)=0$. Hereafter, only the case $\Lambda=0$ will be considered. The $f$ function appearing in Eq. (\ref{theeq}) can be written as
\begin{equation}
\label{newf}
f(r) =\frac{p(r)}{(q^{2}+6Mr)^{3}}, 
\end{equation}
where we have defined the auxiliary function  $p(r)$ as
\begin{equation}
\label{p(r)}
p(r) = 216M^{3}r^{3}+108M^{2}(q^{2}-4M^{2})r^{2}+18Mq^{4}r+q^{6}.
\end{equation}

\begin{figure}
\center{\includegraphics[scale=0.4]{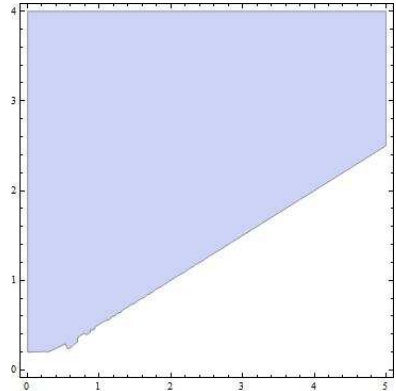}}  \\
\caption{Graphical solution of Eq. (\ref{condition}). The mass $M$ and charge $q$ are plotted in the $y$ and $x$--axis, respectively. }
\label{fig1}
\end{figure}

Clearly the horizons are located at $p(r)=0$. However, in order to have a physically meaningful model, we are interested in the case of three real roots.  
This can be achieved by forming the discriminant from the corresponding polynomial equation and then analysing when it is negative \cite{libroR}.  
That is,  
\begin{eqnarray}
\label{condition}
D &=&  n^{2}+m^{3} < 0,
\end{eqnarray}
where $m$ and $n$ are certain functions of the mass $M$ and charge $q$ \cite{libroR}. When solving for small values of $q$ and $M$, the previous condition 
has a graphic solution given by the shaded region of Fig. (\ref{fig1}).
However, we note that the linear tendency is maintained  for bigger values both of $M$ and $q$.

As Fig. (\ref{fig1}) suggests, there exists a region where the polynomial (\ref{p(r)}) has three real roots. From this general case of three real roots 
we are interested in the particular case of two positive and one negative as it would be desirable to resemble the Reissner-N\"{o}rdstrom case and to avoid 
degeneracies in the horizons.
For this particular case the  actual limit for the mass of the BH would be
\begin{equation}
\label{secondr}
M \geq q
\end{equation}
To get a glimpse of what is happening with this new limit it suffices to use the following properties for the roots of a cubic equation \cite{libroR}. If $x_{1} , x_{2}$ and $x_{3}$ are the roots of an arbitrary cubic equation, then\\
\begin{eqnarray}
 x_{1} + x_{2} + x_{3} = - \frac{b}{a} \nonumber \\
 x_{1} x_{2}  x_{3} = - \frac{d}{a} \nonumber \\
\frac{1}{ x_{1}} + \frac{1}{x_{2}} +\frac{1}{ x_{3}} = - \frac{c}{d} \nonumber \\
\nonumber
\end{eqnarray}
Where in our specific case: $\frac{b}{a} = \frac{q^{2}-4M^{2}}{2M}$,  $ \frac{d}{a} = \frac{q^{6}}{26M^{3}}$ and $ \frac{c}{d} = \frac{18M}{q^{2}}$. If one assumes that $M\gg q$, which is the limit with no pathologies in the sign and existence of the three different roots, then it is straightforward to show that, 
at first order in $q$, 
\begin{equation}
\label{relations}
 x_{1} , x_{2} \approx 0  \hspace{0.2cm} \text{but} \hspace{0.2cm}  x_{1} = - x_{2} \hspace{0.2cm} \text{and} \hspace{0.2cm} x_{3} \approx 2M. \\
\end{equation}
Which is a good symptom of our theory since this case would describe a Schwarzschild BH with a charge perturbing it. Finally, if we use again 
the relations given by Eq. (\ref{relations}), this time relaxing the condition to $\vert q \vert \approx \vert M \vert$ but still having $x_{3} \approx 2M$, 
then the error made to the other roots $x_{1}$ and $x_{2}$ is about $\epsilon \approx \frac{1}{532}$ which is a very small number. From this we conclude that
the limit $M \geq q$ has a solid footing.
 
To summarize, in this discussion we have shown that for any combination of $ M \geq q$, $p(r)$ has two real positive roots which are very close to the horizons
of a perturbed Schwarzschild BH, i.e:  $r_{-} \approx 0$ and $r_{+} \approx 2M$. The perturbation is performed by the charge $q$ and it has the 
additional and very important feature of regularising the geometry.

Therefore, following the previous reasoning, in the next section we are going to work in the regime $M\gg q$, which guarantees two horizons and no 
pathologies since it fulfills the condition given by Eq. (\ref{secondr}). We also note that this regime avoids the case of having a naked singularity 
that would defy the Penrose cosmic censorship conjecture, which is believed to be true \cite{Penrose1965}.
Within this limit,  the two horizons of the regular BH, $r_+$ and $r_-$ are given by
\begin{eqnarray}
\label{horizons}
r_+&=&2M \left[1-\left(\frac{q}{2M}\right)^2\right] +\mathcal{O} \left(\frac{q}{2M}\right)^3\nonumber \\
r_-&=&\frac{|q^3|}{12\sqrt3 M^2} + \mathcal{O}\left(\frac{q}{2M}\right)^3
\end{eqnarray}
(remember that the third root of $f(r)$ is given by $r=-r_-$ but is not of physical interest as it can not be interpreted as a horizon).
\section{Thermodynamics}
Given the expressions for the horizons as expressed by Eqs. (\ref{horizons}), in this section a discussion of the thermodynamics of this regular BH will be 
presented. 

As in the the Reissner-Nordstr\"om case, the surface gravity is given by
\begin{equation}
\kappa=\frac{r_+-r_-}{2r_+^2}.
\end{equation}
%
%
%
Therefore, the temperature is readily obtained using the equation $T = \frac{\kappa}{2 \pi}$. We obtain
\begin{equation}
T=\frac{1}{8\pi M}+\frac{q^2}{32M^3\pi}+\mathcal{O}\left(\frac{q}{2M}\right)^3
\label{Tem}
\end{equation}
and, therefore, using the second principle in the form, $dS= dM \frac{2\pi}{\kappa}$, the entropy of the BH takes the form
\begin{eqnarray}
\label{entropy}
S&=& 4M^{2}\pi-2\pi q^{2}\ln(M)+ \mathcal{O}\left(\frac{q}{2M}\right)^3\nonumber \\
&=& \frac{\mathcal{A}}{4}-\pi q^2\ln\left(\mathcal{A}\right) +S_{0}+ \mathcal{O}\left(\frac{q}{2M}\right)^3,
\end{eqnarray}
where $\mathcal{A}$ is the Schwarzschild BH area and $S_{0} =\nobreak 2 \pi q^{2}\ln\left(\frac{1}{2 \sqrt{ \pi }}\right)$ is a constant.

In addition, an interesting result obtained from this particular solution is the emergence of a remnant mass, which can be obtained by calculating the 
heat capacity, defined as $C=\frac{dM}{dT}$. After some calculations we get
\begin{equation}
C=-8 \pi M^2 + 6 \pi q^2 + \mathcal{O}\left(\frac{q}{2M}\right)^3.
\label{HC}
\end{equation}\\
%

Therefore, as the BH evaporation stops when $C(M_r)=\nobreak0$, a remnant mass can be obtained. From  Eq. (\ref{HC}), this mass is shown to be given by
\begin{equation}
M_r=\sqrt{\frac{3}{4}}\,q.
\end{equation}
%


\section{Non--linear electrodynamics realization of a generalized uncertainty principle?}

There is an intriguing similitude between the entropy of the regular BH, given by Eq. (\ref{entropy}), and that of a Schwarzschild BH with 
quantum corrections. Specifically, different approaches to quantum gravity predict corrections to the Bekenstein--Hawking entropy in
the form 
\cite{Kaul2000,Medved2004,Camelia2004,Meissner2004,Das2002,Domagala2004,Chatterjee2004,Akbar2004,Myung2004,Chatterjee2005,Aros2010,Sen2013,Aros2013}
\begin{equation}
\label{eqfirst}
S=\frac{\mathcal{A}}{4}+c_{0}\ln \left(\frac{\mathcal{A}}{4} \right)+\sum_{n=1}^{\infty}c_{n}\left(\frac{\mathcal{A}}{4} \right)^{-n},
\end{equation}
where the $c_{n}$ coefficients are parameters which depend on the specific model we are considering. 

Among all the models employed to extract
conclusion from the merging between gravity and the quantum world, a minimal length scale is thought to be a essential ingredient \cite{Sabine2013}. 
This minimum length can be implemented, for example, as a generalized uncertainty principle (GUP), which has been very recently derived from
first principles \cite{inPRESS}. This GUP can be implemented using deformed commutation relations which
include terms linear (or quadratic) in the particle momenta as
\begin{equation}
x_{i}=x_{0i}\; ;  \; p_{i}=p_{0i}\,\left(1-\alpha p_{0}+ 2 \alpha^{2} p_{0}^{2}\right),
\end{equation}
where $\left[x_{0i},p_{0j}\right]=i \hbar \delta_{ij}$ and $p_{0}^{2}=p_{0j}p_{0j}$ ($j=1,2,3$) and
$\alpha$ is a dimensionless constant.
These GUP effects have been used to compute the BH entropy \cite{Medved2004,Adler2001,Camelia2006,Majumder2011,Majumder2013,Bargueno2015}, 
which is given by
\begin{equation}
\label{GUP}
S= \frac{\mathcal{A}}{4 }+\frac{\sqrt{\pi} \alpha}{4}\sqrt{\frac{\mathcal{A}}{4}}-\frac{\pi \alpha^{2}}{64}\ln 
\left(\frac{\mathcal{A}}{4 } \right)+ ...
\end{equation}

Therefore, given the similarity between Eqs. (\ref{entropy}) and (\ref{GUP}), both the charge and the GUP parameter can be related by
\begin{equation}
q\leftrightarrow \frac{\alpha}{8},
\end{equation}
where only a quadratic GUP is considered \footnote{Remember that we are taking $\hbar=c=G=1$ and, therefore, $m_{p}=l_{p}=1$.}. Moreover, 
the existence of a remnant in the charged case here considered is in complete agreement with similar
conclusions obtained within the quadratic GUP formalism \cite{Banerjee2010,Dutta2014}. Thus, a quadratic GUP can be clearly realized by the NLED model
here employed which, in this sense, acquires more importance.

\section{Conclusion}

In this work, we have managed to derive a regular black hole solution which solves the Einstein equations when certain non--linear electrodynamics model 
is invoked. This model here considered satisfies the weak energy condition. Moreover, although the matter content is shown to be similar, for small
fields, to the Ay\'on--Beato and Garc\'{\i}a solution, the electric field here considered has a very simple mathematical expression, in contrast to
other cases. We have computed the thermodynamic properties of this regular black hole by a careful analysis of the horizons and we have found that the
usual Bekenstein--Hawking entropy gets corrected by a logarithmic term. Therefore, in this sense our model realizes some quantum gravity predictions
which add a logarithmic correction to the black hole entropy. In particular, we have established some similitudes between our model and a quadratic generalized
uncertainty principle by showing that the charge plays the role of the $\alpha$--parameter which enters into the later. This similitude has been confirmed
by the existence of a remnant, which prevents complete evaporation, in complete agreement with the quadratic generalized uncertainty principle case.

Finally, we would like to point out that the following cases give place to asymptotically flat regular solutions (not shown here):
\\
\begin{itemize}
\item $n=2 \gamma$ ($\gamma$ is a positive natural number $\ge 2$) when $a_{n}=\frac{2q^2}{3(n+1)M}$
\item $n=2 \gamma+1$ ($\gamma$ is a positive natural number $\ge1$) when $a_{n}=2\,\frac{2q^2}{3(n+1)M}$
\end{itemize}
We note that, although it has been shown that the $n=3$ case represents a regular black hole, other regular solutions with different $n$, together with their energy conditions, remain to be studied.\\\\

The authors acknowledge support from the Faculty of Science and Vicerrector\'{\i}a de Investigaciones of
Universidad de los Andes, Bogot\'a, Colombia, under FAPA project ``Enredamiento, gravitaci\'on y sistemas cu\'anticos".


\begin{thebibliography}{}
\bibitem{Penrose1965}R. Penrose, Phys. Rev. Lett. {\bf14}, 57 (1965).
\bibitem{book}S. W. Hawking and G. F. R. Ellis,  The Large Scale Structure of space--time, Cambridge University Press, Cambridge (1979).
\bibitem{Senovilla}J. M. Senovilla, arXiv:0605007.
\bibitem{HE}W. Heisenberg and H. Euler, Z. Phys. {\bf 120}, 714 (1936).
\bibitem{Sch}J. Schwinger, Phys. Rev. {\bf 82}, 664 (1951).
\bibitem{BI}M. Born and L. Infeld, Proc. Roy. Soc. London A {\bf 144}, 425 (1934); ibid. {\bf 143}, 410 (1934) and {\bf 147}, 522 (1934).
\bibitem{Bia}Z. Bialynicka--Birula, I. Bialynicki--Birula, Phys. Rev. D {\bf 2}, 2341 (1970).
\bibitem{PLB1985}E. S. Fradkin and A. A. Tseytlin, Phys. Lett B. {\bf 163}, 123 (1985).
\bibitem{NPB1997}A. A. Tseytlin, Nucl. Phys. B {\bf 501}, 41 (1997).
\bibitem{Garcia1984}A. Garc\'{\i}a D., H. Salazar I. and J. F. Pleba\'nski, Nuovo Cimento Ser. B {\bf 84}, 65 (1984).
\bibitem{Breton2003}N. Breton, Phys. Rev. D {\bf 67}, 124004 (2003).
\bibitem{Hendi2013}S. Hendi, Ann. Physics {\bf 333}, 282 (2013).
\bibitem{Ayon1998}E. Ayon--Beato and A. Garcia, Phys. Rev. Lett. {\bf 80}, 5056 (1998).
\bibitem{Ayon1999a}E. Ayon--Beato, A. Garc\'{\i}a, Phys. Lett. B {\bf 464}, 25 (1999).
\bibitem{Ayon1999b}E. Ayon--Beato, A. Garc\'{\i}a, Gen. Relativ. Gravit. {\bf 31}, 629 (1999).
\bibitem{Ayon2000}E. Ayon--Beato, A. Garc\'{\i}a, Phys. Lett. B {\bf 493}, 149 (2000).
\bibitem{Ayon2005}E. Ayon--Beato, A. Garc\'{\i}a, Gen. Relativ. Gravit. {\bf 37}, 635 (2005).
\bibitem{Dym2004} I. Dymnikova, Class. Quantum Gravity, {\bf 21}, 4417 (2004).
\bibitem{Baldo2000}F. Baldovin, M. Novello, S.E. Perez Bergliaffa and J. Salim, Classical Quantum Gravity {\bf 17}, 3265 (2000).
\bibitem{Bronni2001}K. A. Bronnikov, Phys. Rev. D {\bf 63}, 044005 (2001).
\bibitem{Bardeen1968}J.M. Bardeen, Proceedings of GR5, Tbilisi, USSR, p. 174 (1968).
\bibitem{Yajima2001}H. Yajima and T. Tamaki, Phys. Rev. D {\bf 63}, 064007 (2001).
\bibitem{Hassaine2008}M. Hassaine and C. Martinez, Classical Quantum Gravity {\bf 25}, 195023 (2008).
\bibitem{Zas2010} O. B. Zaslavskii, Phys. Lett. B, {\bf 688}, 278 (2010).
\bibitem{Ansoldi2007}S. Ansoldi, P. Nicolini, A. Smailagic and E. Spallucci, Phys. Lett. B {\bf 645}, 261 (2007).
\bibitem{Balart2014a}L. Balart and E. C. Vagenas, Phys. Lett. B {\bf 730}, 14 (2014).
\bibitem{Balart2014b}L. Balart and E. C. Vagenas, Phys. Rev. D {\bf 90}, 124045 (2014).
\bibitem{Contreras2016}E. Contreras, F. D. Villalba and P. Bargue\~no, EPL (accepted, 2016).
\bibitem{Cherubini2002}S. Cherubini. D. Bini, S. Capozziello and R. Ruffini, Int. Journal Mod. Phys. D {\bf 11}, 827 (2002).
\bibitem{Einstein1935}A. Einstein and N. Rosen,  Phys. Rev. {\bf 48}, 73 (1935).
\bibitem{Salazar1987}H. Salazar I., A. Garc\'{\i}a D. and J. Pleba\'nski, J. Math. Phys. {\bf 28}, 2171 (1987).
\bibitem{Bronnikov2001} K. A. Bronnikov, Phys. Rev. D {\bf 63}, 044005 (2001).
\bibitem{Graves1960}J. G. Graves, D. Brill, Physical Review. {\bf120}, 4 (1960).
\bibitem{libroR} I.N. Bronshtein, K.A Semendyayev, G.Musiol, H.Muehlig, Handbook of Mathematics, Springer, Leipzig (2005).
\bibitem{Kaul2000} R. K. Kaul and P. Majumdar, Phys. Rev. Lett {\bf 84 }, 5255 (2000).
\bibitem{Medved2004}A. J. Medved and E. C. Vagenas, Phys. Rev.  {\bf D70}, 124021 (2004).
\bibitem{Camelia2004}G. A. Camelia, M. Arzano and A. Procaccini, Phys. Red. D {\bf 70}, 107501 (2004).
\bibitem{Meissner2004}K. A. Meissner, Class. Quant. Grav. {\bf 21}, 5245 (2004).
\bibitem{Das2002} S. Das, P. Majumdar and R. K. Bhaduri, Class. Quant. Grav. {\bf 19}, 2355 (2002).
\bibitem{Domagala2004}M. Domagala and J. Lewandowski, Class. Quant. Grav. {\bf 21}, 5233 (2004).
\bibitem{Chatterjee2004} A. Chatterjee and P. Majumdar, Phys. Rev. Lett. {\bf 92}, 141301 (2004).
\bibitem{Akbar2004}M. M. Akbar and S. Das, Class. Quant. Grav. {\bf 21}, 1383 (2004).
\bibitem{Myung2004}Y. S. Myung, Phys. Lett.  {\bf B579}, 205 (2004).
\bibitem{Chatterjee2005}A. Chatterjee and P. Majumdar, Phys. Rev.  {\bf D71}, 024003 (2005).
\bibitem{Aros2010}R. Aros, D.E. D\'{\i}az and A. Montecinos, JHEP, 012 (2010).
\bibitem{Sen2013}A. Sen, JHEP, 156 (2013).
\bibitem{Aros2013}R. Aros, D.E. D\'{\i}az and A. Montecinos, Phys. Rev. D {\bf 88}, 104024 (2013).
\bibitem{Sabine2013}S. Hossenfelder, Living Rev. Relativity, {\bf 16}, 2 (2013).
\bibitem{inPRESS}R. N. Costa Filho, J. P. M. Braga, J. H. S. Lira and J. S. Andrade Jr., Phys. Lett B, {\bf 755}, 367 (2016).
\bibitem{Adler2001}R. J. Adler, P. Chen and D. I. Santiago, Gen. Relativit. Gravit. {\bf 33}, 2101 (2001).
\bibitem{Camelia2006}G. Amelino--Camelia, M. Arzano, Y. Ling and G. Mandanici, Class. Quant. Grav. {\bf 23}, 2585 (2006).
\bibitem{Majumder2011}B. Majumder, Phys. Lett  {\bf B703}, 402 (2011).
\bibitem{Majumder2013}B. Majumder, Gen. Relativit. Gravit. {\bf 45}, 2403 (2013).
\bibitem{Bargueno2015}P. Bargue\~no and E. C. Vagenas, Phys. Lett. B {\bf 742}, 15 (2015).
\bibitem{Banerjee2010} R. Banerjee and S. Ghosh, Phys. Lett. {\bf B688}, 224 (2010).
\bibitem{Dutta2014}A. Dutta and S. Gangopadhyay, Gen. Rel. Gravit. {\bf 46}, 1747 (2014).
\end{thebibliography}
\end{document}